\documentclass[useAMS,usenatbib]{mn2e}
\usepackage{color}
\usepackage{graphicx}
\usepackage{dcolumn}
\usepackage{bm}
\usepackage{amssymb}
\usepackage{latexsym}
\usepackage[T1]{fontenc}
\usepackage{aecompl} 

\newcommand{\hMsun}{{\ifmmode{h^{-1}{\rm
        {M_{\odot}}}}\else{$h^{-1}{\rm{M_{\odot}}}$~}\fi}} 
\newcommand{\hMpc}{{\ifmmode{h^{-1}{\rm Mpc}}\else{$h^{-1}$Mpc }\fi}}

\begin{document}

\title[Constraints using the Alcock-Paczynski effect]{Cosmological constraints from the redshift dependence of the Alcock-Paczynski test and volume effect: galaxy two-point correlation function}

\author[Xiao-Dong~Li, Changbom~Park, Cristiano G. Sabiu and Juhan Kim]
{ Xiao-Dong Li$^{\dagger}$, Changbom Park$^{1}$, Cristiano G. Sabiu$^{2,1,\star}$, Juhan Kim$^{3,1}$\\
$^1$School of Physics, Korea Institute for Advanced Study, 85 Heogi-ro, Dongdaemun-gu, Seoul 130-722, Korea\\
$^2$Korea Astronomy and Space Science Institute, 776, Daedeokdae-ro, Yuseong-gu, Daejeon, 305-348, Korea\\
$^3$Center for Advanced Computation, Korea Institute for Advanced Study, 85 Hoegi-ro, Dongdaemun-gu, Seoul 130-722, Korea\\
$^{\dagger}$xiaodongli@kias.re.kr\\
$\star$Corresponding Author: csabiu@kasi.re.kr}

%\date{Accepted 1988 December 15. Received 1988 December 14; in original form 1988 October 11}

% \begin{keywords}
% methods: data analysis -- methods: statistical -- Galaxies:
% kinematics and dynamics -- Cosmology: observations -- large-scale
% structure of universe
% \end{keywords}

\pagerange{\pageref{firstpage}--\pageref{lastpage}} \pubyear{2002}

\maketitle

\label{firstpage}

\begin{abstract}
We propose a method using the redshift dependence of the Alcock-Paczynski (AP) test and volume effect to measure the cosmic expansion history.
The galaxy two-point correlation function as a function of angle, $\xi(\mu)$, is measured at different redshifts.
Assuming an incorrect cosmological model to convert galaxy redshifts to distances,
the shape of $\xi(\mu)$ appears anisotropic due to the AP effect,
and the amplitude shifted by the change in comoving volume.
Due to the redshift dependence of the AP and volume effect,
both the shape and amplitude of $\xi(\mu)$ exhibit redshift dependence.
Similar to \cite{Li2014}, we find the redshift-space distortions (RSD) caused by galaxy peculiar velocities,
although significantly distorts $\xi(\mu)$,
exhibit much less redshift evolution compared to the AP and volume effects.
By focusing on the redshift dependence of $\xi(\mu)$,
we can correctly recover the cosmological parameters despite the contamination of RSD.
The method is tested by using the Horizon Run 3 N-body simulation,
from which we made a series of $1/8$-sky mock surveys
having 8 million physically self-bound halos and sampled to have roughly a uniform number density in $z=0-1.5$.
We find the AP effect results in tight, unbiased constraints on the density parameter and dark energy equation of state,
with 68.3\% CL intervals $\delta \Omega_m\sim0.03$ and $\delta w\sim0.1$,
and the volume effect leads to much tighter constraints of $\delta \Omega_m\sim0.007$ and $\delta w\sim0.035$.
\end{abstract}

% \begin{keywords}
% circumstellar matter -- infrared: stars.
% \end{keywords}

\section{Introduction}

Since the discovery of cosmic acceleration \citep{Riess1998,Perl1999}, 
there have been many models proposed to explain it. These include modifications to general relativity (GR) 
and additional energy components to the constituents of the Universe. 
The simplest and currently favored solution is the additional field called dark energy which has negative pressure and is spatially homogeneous.
%The simplest explanation for dark energy is that it is a cosmological constant or vacuum energy, with constant equation of state (EoS) $w = -1$.

To understand the nature of dark energy, 
we need to make precise measurements of the cosmic expansion history using cosmological observables. 
Two such observables are the angular diameter distance $D_A$ and the Hubble factor, $H$. 
If these can be measured at various redshifts, then tight constraints can be placed on the parameters of the cosmological model,
e.g., the matter density $\Omega_m$ and the equation of state (EoS) of dark energy, $w$.

Assuming an incorrect cosmological model for the coordinate transformation from redshift space to comoving space leaves residual geometric distortions. 
These distortion, known as the redshift-space distortions (RSD), are induced by the fact that distances along and perpendicular to the line of sight are fundamentally different.
Measuring the ratio of galaxy clustering in the radial and transverse directions, provides a probe of the Alcock-Paczynski (AP) effect \citep{AP1979}.

There have been several methods proposed that apply the AP test to the large scale structure
by measuring the clustering of galaxies \citep{Ballinger1996,Matsubara1996},
symmetry properties of galaxy pairs \citep{Marinoni2010,Jennings2011,BB2012}, 
and cosmic voids \citep{Ryden1995,LavausWandelt1995,Sutter2014}.
Among them, the method of galaxy clustering has been widely used to constrain cosmological parameters 
\citep{Outram2004,Blake2011,ChuangWang2012,Reid2012,Beutler2013,Linder2013,2014arXiv1407.2257S, Jeong2014,2014ApJ...781...96L}.

%However, one advantage of our analysis is that it is significantly model independent and does not assuming LCDM or any other dark energy model.
%Instead we directly fit for the angular distance $D_A$, Hubble parameter $H$.
The main caveat of this method is that, because the radial distances of galaxies are inferred from redshifts, 
AP tests are inevitably limited by RSD,
which leads to apparent anisotropy even if the adopted cosmology is correct \citep{Ballinger1996}.
%This effect must be then accurately modeled for the 2-point statistics of galaxy clustering.
%To avoid the contamination induced by RSD,
In \cite{Li2014} we proposed a new method utilizing the redshift dependence of AP effect to overcome the RSD problem, 
which uses the isotropy of the galaxy density gradient field. 
We found that the redshift dependence of the anisotropy created by RSD is much less significant compared with the anisotropy caused by AP.
Thus we focused on the redshift dependence of the galaxy density gradient field,
which is less affected by RSD, but still sensitive to cosmological parameters.

The two-point correlation analysis is the most widely used method to study the large scale clusterings of galaxies.
So, in this paper we revisit the topic of \cite{Li2014} using the galaxy two-point correlation function (2pCF).
By investigating the redshift dependence of anisotropic galaxy clustering we can measure the AP effect despite of contamination from RSD.
Moreover, when the redshift evolution of galaxy bias can be reliably modelled,
we can measure the redshift evolution of volume effect from the amplitude of 2pCF.
%Except the AP distortion of the shapes of structures, 
The change of the comoving volume size is another consequence of a wrongly adopted cosmology,
which has motivated methods constraining cosmological parameters from number counting of 
galaxy clusters \citep{PS1974,VL1996} and topology \citep{topology}.

The outline of this paper proceeds as follows. 
In \S 2 we briefly review the nature and consequences of the AP effect and volume changes when performing coordinate transforms in a cosmological context. 
In \S 3 we describe the N-body simulations and mock galaxy catalogues that are used to test our methodology.
In \S 4 and \S 5, we describe our new analysis method for quantifying the anisotropic clustering as well as proposing a way to deal with the RSD that are convolved with the AP distortion. 
Here we also present results of our optimised estimator. 
We conclude in \S 6.

\section{AP and Volume Effect in Wrongly Assumed Cosmologies}
\label{sec:APeffect}

\begin{figure*}
   \centering{
   \includegraphics[height=6cm]{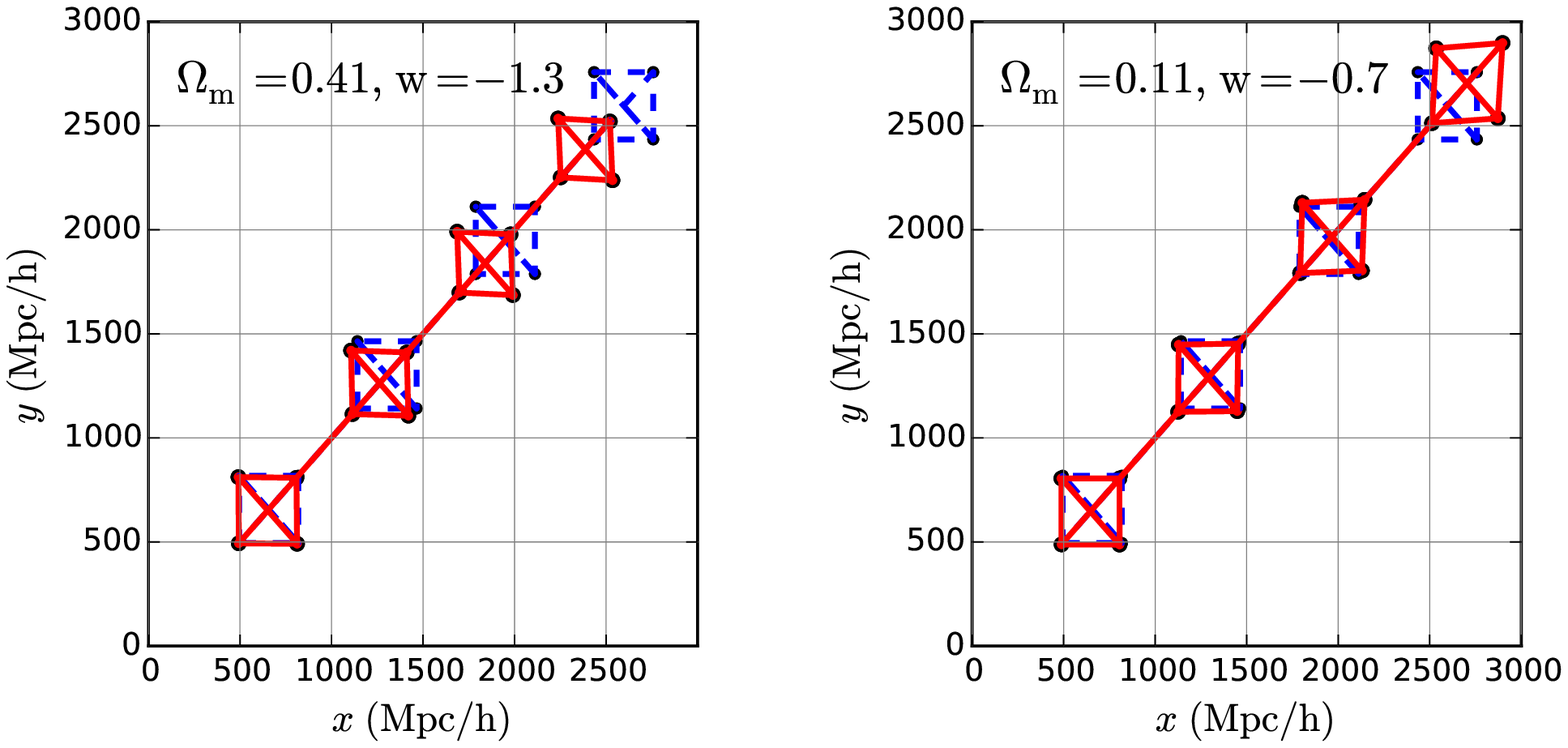}
   \includegraphics[height=6cm]{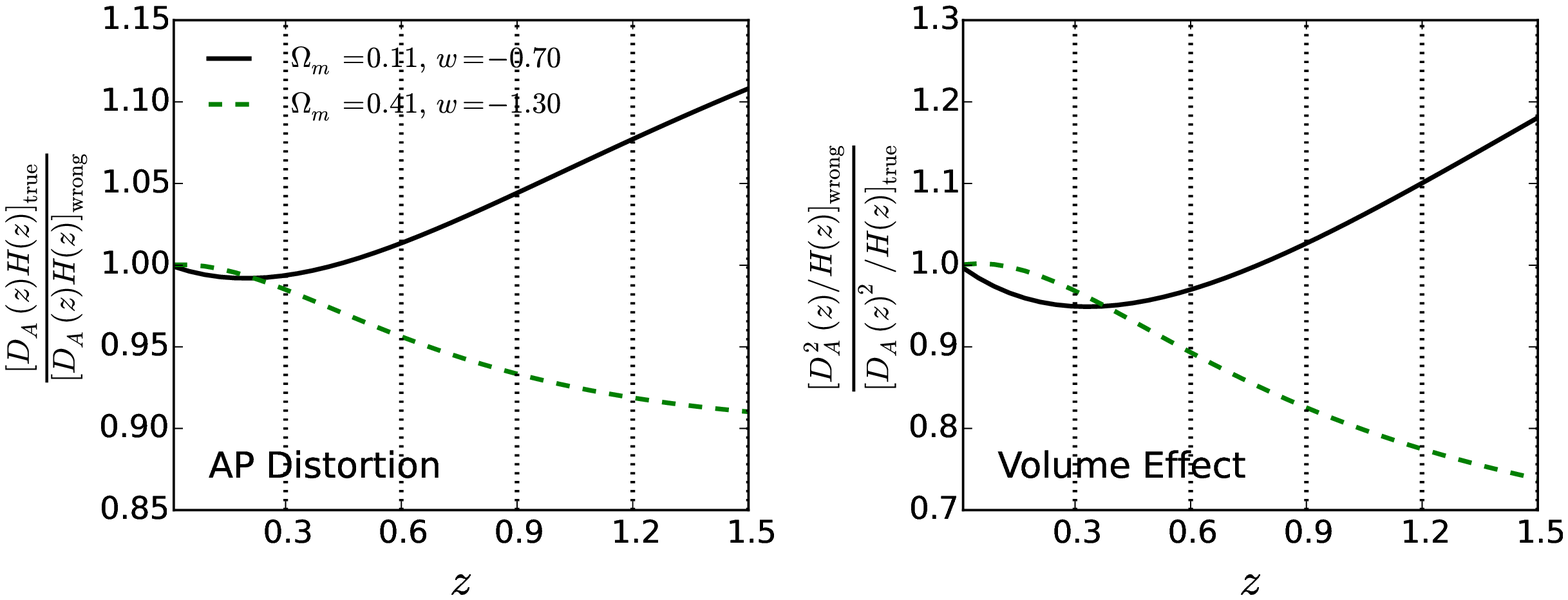}
   }
   \caption{\label{fig_xy}
   %Apparent distortion of objects in four wrongly assumed cosmologies, assuming a true cosmology of .
   The redshift dependence of AP and volume effect in two wrongly assumed cosmologies $\Omega_m=0.41$, $w=-1.3$ and $\Omega_m=0.11$, $w=-0.7$,
   assuming a true cosmology of $\Omega_m=0.26$, $w=-1$.
   Upper panel shows the apparent distortion of four perfect squares,
   measured by an observer located at the origin.
   The apparently distorted shapes are plotted in red solid lines.
   The underlying true shapes are plotted in blue dashed lines.
   Lower panel shows the evolution of Equations (\ref{eq:stretch}) and (\ref{eq:volume}).
   In our mock surveys we split the samples at $z=0.3$, 0.6, 0.9 and 1.2, as marked by the vertical lines.
   %measured by an observer located at the origin.
   %For reference, blue dashed squares show their shapes and positions in the true cosmology.
   }
\end{figure*}

%Considering we are Let us consider an object in the universe with certain size and shape.
%Its observed redshift span $\Delta z$ and angular size $\Delta \theta$ are related with the comoving sizes by
In this section we briefly introduce the AP and volume effect in wrongly assumed cosmologies.
A more detailed description of the AP effect has been provided in \cite{Li2014}.

Suppose that we are probing the shape and volume of an object in the Universe.
We measure its redshift span $\Delta z$ and angular size $\Delta \theta$,
then compute its sizes in the radial and transverse directions from the relations of 
\begin{equation}\label{eq:distance}
\Delta r_{\parallel} = \frac{c}{H(z)}\Delta z,\ \ \Delta r_{\bot}=(1+z)D_A(z)\Delta \theta,
\end{equation}
where $H$ is the Hubble parameter, $D_A$ is the angular diameter distance.
In the particular case of a flat universe with constant dark energy EoS, they take the forms of
\begin{eqnarray}\label{eq:HDA}
& &H(z) = H_0\sqrt{\Omega_ma^{-3}+(1-\Omega_m)a^{-3(1+w)}},\nonumber\\
& &D_A(z) = \frac{1}{1+z}r(z)=\frac{1}{1+z}\int_0^z \frac{dz^\prime}{H(z^\prime)},
\end{eqnarray}
where $a=1/(1+z)$ is the cosmic scale factor,
$H_0$ is the present value of Hubble parameter and $r(z)$ is the comoving distance.

In case we adopted a wrong set of cosmological parameters in Equation (\ref{eq:distance},\ref{eq:HDA}),
the inferred $\Delta r_{\parallel}$ and $\Delta r_{\bot}$ are wrong,
resulting in distorted shape (AP effect) and wrongly estimated volume (volume effect).
The degree of variations in shape and volume can be described by the following quantities
\begin{equation}\label{eq:stretch}
 %{\rm Degree\ of\ LOS\ stretch\equiv}
 %{\rm rat}_{\rm LOS\ stretch}\equiv
 \frac{[\Delta r_{\parallel}/\Delta r_{\perp}]_{\rm wrong}}{[\Delta r_{\parallel}/\Delta r_{\perp}]_{\rm true}} =
  \frac{[D_A(z)H(z)]_{\rm true}}{[D_A(z)H(z)]_{\rm wrong}} 
\end{equation}
\begin{equation}\label{eq:volume}
 %{\rm Degree\ of\ volume\ magnification \equiv}
 %{\rm rat}_{\rm volume\ mag}\equiv
 \frac{[\Delta r_{\parallel}(\Delta r_{\perp})^{2}]_{\rm wrong}}{[\Delta r_{\parallel}(\Delta r_{\perp})^{2}]_{\rm true}}
 = \frac{{\rm Volume}_{\rm wrong}}{{\rm Volume}_{\rm true}}
 = \frac{[D_A(z)^2/H(z)]_{\rm wrong}}{[D_A(z)^2 / H(z)]_{\rm true}},
\end{equation}
where ``true'' and ``wrong'' denote the values of quantities in the true cosmology and wrongly assumed cosmology.
From the AP and volume effects, we can constrain  $D_A(z)H(z)$ and $D_A(z)^2 / H(z)$, respectively.

The AP and volume effect due to wrongly assumed cosmological parameters are shown in the upper panel of Figure \ref{fig_xy}.
Suppose that the true cosmology is a flat $\Lambda$CDM with present density parameter $\Omega_m=0.26$
and standard dark energy EoS $w=-1$.
If we were to distribute four perfect squares at various distances from 500 Mpc/h to 3,000 Mpc/h,
and an observer located at the origin were to measure their redshifts and compute their positions and shapes 
using redshift-distance relations of two incorrect cosmologies:
\begin{enumerate}
 \item $\Omega_m=0.41$, $w=-1.3$,
 \item $\Omega_m=0.11$, $w=-0.7$,
\end{enumerate}
the shapes of the squares appear distorted (AP effect),
and their volumes are changed (volume effect).
In the cosmological model (ii) with $\Omega_m=0.11$, $w=-0.7$, we see a stretch of the shape in the line of sight (LOS) direction (hereafter ``LOS shape stretch'')
and magnification of the volume (hereafter ``volume magnification''), 
while in the model with $\Omega_m=0.41$, $w=-1.3$, we see opposite effects of LOS shape compression and volume shrinkage.

In the lower panel of Figure \ref{fig_xy}, we plot the degree of variations in shape and volume as functions of redshift. 
In cosmology (i), both quantities have values less than 1, 
indicating LOS shape compression and volume shrink.
The effect in cosmology (ii) is slightly more subtle. At low redshift, the effect of dark energy is important, 
and there is LOS shape compression and volume reduction due to the quintessence like dark energy EoS.
However, at higher redshift the role of dark matter is more important, and we see LOS shape stretch and volume magnification due to the small $\Omega_m$.

More importantly, Figure \ref{fig_xy} highlights the redshift dependence of the AP and volume effects. 
For example, in the cosmology with $\Omega_m=0.41$, $w=-1.3$, 
both the LOS shape stretch and volume magnification become more significant with increasing redshift.
In the cosmology with $\Omega_m=0.11$, $w=-0.7$,
not only do the magnitudes of the effects evolve with redshift,
but there is also a turnover from LOS shape compression and volume shrink at lower redshift to LOS shape stretch and volume magnification at higher redshift.

\section{Mock Data}
\label{sec:mocks}
We test the methodology using mock surveys constructed from one of the Horizon Run simulations, HR3.
The Horizon Run simulations are a suite of large volume N-body simulations that have resolutions and volumes capable of reproducing the observational statistics of many current major redshift surveys like the Sloan Digital Sky Survey's (SDSS) Baryon Oscillation Spectroscopic Survey (BOSS) etc \citep{park 2005,2009ApJ...701.1547K,horizonrun}.
HR3 adopts a flat-space $\Lambda$CDM cosmology with the WMAP 5 year parameters
$\Omega_{m}=0.26$, $H_{0}=72{\rm km/s/Mpc}$, $n_{s}=0.96$ and $\sigma_8=0.79$ \citep[]{komatsu 2011}.
The simulation was made in a cube of volume $(10.815 {~ h^{-1}} {\rm{Gpc}})^3$
using $7120^3$ particles with particle mass of $1.25\times 10^{11}$\hMsun.

The simulations were integrated from $z=27$ and reached $z=0$ after making $N_{\rm step}=600$ timesteps.
Dark matter halos were identified using the Friend-of-Friend algorithm with the linking length of 0.2 times the mean particle separation.
Then the physically self-bound (PSB) subhalos that are gravitationally self-bound and tidally stable are identified \citep{kim and park 2006}.
The PSB halo finder is a group finding algorithm which can efficiently identify halos located even in crowded regions. 
This method combines two physical criteria such as the tidal radius of a halo and the total energy of each particle to find member particles.
%This provides a substantial increase in the similarity between simulation and observational data,
%as these dark matter subhalos are sites for galaxy formation.
%LRG (Xiao: is it OK to change all the ``LRG'' to ``galaxy''?) formation.
%To simulate the SDSS survey,
%HR3 places 27 observers evenly within its cubical volume and allows
%each observer to see out to a redshift of 0.7, creating 27
%independent, non-overlapping spherical regions.

An all-sky, very deep light cone survey reaching redshift $z=4.3$ was made by placing an observer located at the center of the box.
The comoving positions and velocities of all CDM particles are saved as they cross the past light cone
and PSB subhalos are identified from this particle data.

To match the observations of recent LRG surveys \citep{gott 2008, gott 2009, choi 2010},
a volume-limited sample of halos with constant number density of $3 \times 10^{-4} (h^{-1}{\rm Mpc})^{-3}$
is selected by imposing a minimum halo mass limit varying along with redshift, 
from $6\times 10^{12}$\hMsun at high redshift to $1.5\times 10^{13}$\hMsun at low redshift.
To do this, the whole sample is split into many thin shells whose radial spans are 30 Mpc/h. 
In each shell, a minimal mass cut is set such that the number density is $\bar n=3\times10^{-4}(h^{-1}{\rm Mpc})^{-3}$. 
These thin shells are then merged together to build up the whole sample. 
Note that all these steps are done before converting the sample to other cosmologies, so the constant number density of halos is only true in the correct cosmology.
%The light cone survey sample consists of subhalos at different redshifts, 
%and thus their redshift dependence on velocities and evolution of clustering are automatically included. 
%The peculiar velocity of the subhalo is set to that of the most-bound particle in that subhalo.
The group velocity of each PSB dark matter halo (or our mock ``galaxy'') 
is used to perturb the redshift of the halo by 
\begin{equation}
\Delta z = (1+z) \frac{v_{{\rm LOS}}}{c},
\end{equation}
where $v_{\rm LOS}$ is the LOS component of the halo proper velocity.

We divide the all-sky survey sample into eight equal sky area subsamples and impose the redshift range $z=0-1.5$. 
This mock data will be relevant for future galaxy spectroscopic surveys \citep[e.g. DESI ][]{2013arXiv1308.0847L}.

% The incorporation of RSD is a crucial point of this method.
% The redshift evolution of RSD effect may depend on the halo finder, the definition of halo velocities, 
% and the properties of the sample (e.g., number density, mass cut). 
% Investigation of these issues are helpful to improve our method.
%We further impose a minimal distance cut of $r>500$ Mpc/h (or equivalently $z>0.17$),
%which is equal to that of the BOSS LOWZ sample. The BOSS LOWZ sample is usually restricted to
%$z>0.15$ where the galaxy number density is more or less uniform \citep{Tojeiro2011,Tojeiro2012,Parejko2013}.
%since low redshift galaxies carry little cosmological information while their positions are severely distorted by RSD.

\section{Methodology}

We probe the effects discussed in \S\ref{sec:APeffect} using the 2pCF. The 2pCF is a mature statistic in cosmology and its optimal estimation considers minimal variance while dealing with complicated masks and selection functions. The most commonly adopted formulation is that of the Landy-Szalay 
estimator~\citep{1993ApJ...412...64L},
\begin{equation}
\xi(s,\mu)=\frac{DD-2DR+RR}{RR}\ ,
\end{equation}
where $DD$ is the number of galaxy--galaxy pairs, $DR$ the number of galaxy-random pairs, and $RR$ is the number of random--random pairs, all separated by a distance defined by 
$s\pm\Delta s$ and $\mu\pm\Delta\mu$, where $s$ is the distance between the pair and $\mu=\cos(\theta)$, with $\theta$ being the angle between the line joining the pair of galaxies and the LOS direction. This statistic therefore captures the radial anisotropy of the clustering signal.

The random point catalogue constitutes an unclustered but observationally representative sample of our mock surveys. 
To reduce the statistical variance of the estimator we use 15 times as many randoms as we have galaxies.

\begin{figure*}
   \centering{
  \includegraphics[height=11cm]{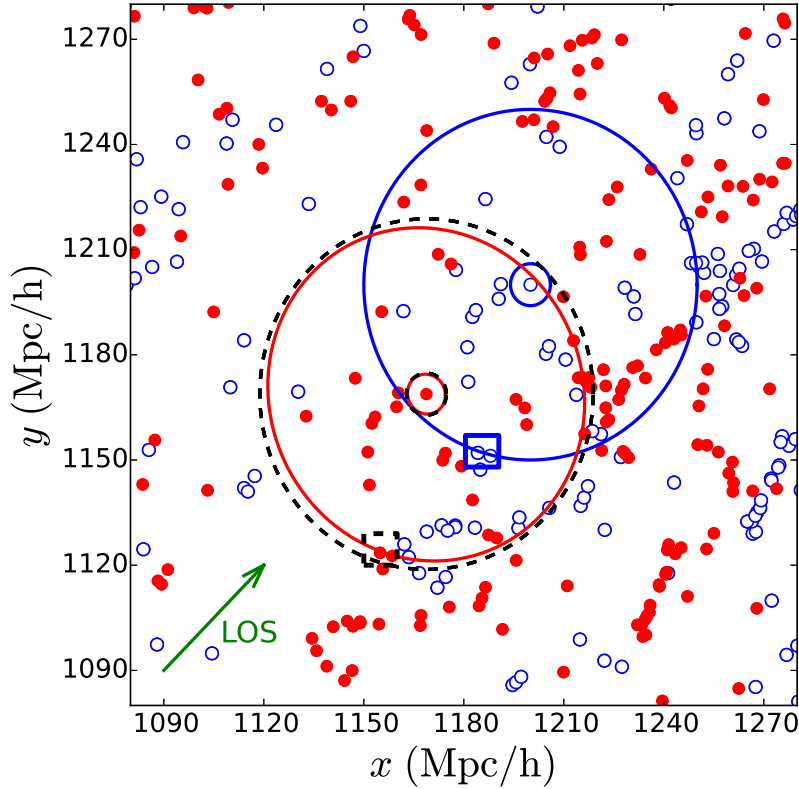}
  }
  \caption{\label{fig_Scatter}
  The influence of the AP and volume effect on the 2pCF statistics, in the wrongly assumed cosmology $\Omega_m=0.41$, $w=-1.3$.
  Scatter points are a 200 Mpc/h $\times$ 200 Mpc/h slice of halos drawn from HR3, with thickness 10 Mpc/h and mean redshift $z=0.43$.
  Their positions are measured by an observer located at the origin, with the LOS direction marked by the green arrow.
  Blue empty circles are their true positions in the correct cosmology, 
  while red filled circles are the apparent positions in the wrong cosmology.
  Blue solid circles are the ``2-point CF rings'' (i.e. the geometrical bins used to evaluate the clustering) 
  with radii 6 and 50 Mpc/h, centered on a galaxy with location $x,y$=1200 Mpc/h.
  Red solid circles are the ``exactly mapped'' rings (bins) in the incorrect cosmology, with its size shrunk and shape distorted.
  Black dashed circles are the actually adopted rings when we measure the 2pCF in the wrong cosmology.
  We expect the result of 2pCF statistics to be affected in three ways.
  First, the shape of structures should appear compressed in the radial direction, introducing anisotropy in $\xi$.
  Secondly, the sizes of structures should appear shrunk due to the volume effect, which shifts the amplitude of $\xi$ .
  Finally, as another consequence of volume effect, structures on larger scales enter the statistics.
  For example, halos within blue solid box are not considered in the correct cosmology,
   but will be considered in the wrong cosmology, as shown by the black dashed box.
   This also shifts the amplitude of $\xi$.
  }
\end{figure*}

To probe the anisotropy, $\xi$ is measured at different directions, 
%In each direction, $\xi(s,\mu_i)$ is measured as a function of scale $s$,
and integrated over the interval $\Delta s = s_{\rm max} - s_{\rm min}$.
We evaluate
\begin{equation}
\xi_{\Delta s} (\mu) \equiv \int_{s_{\rm min}}^{s_{\rm max}} \xi (s,\mu)\ ds.%,\ \ \ {\rm at\ particular\ binned\ direction\ }\mu_i
\end{equation}
%We choose $s_{min}=10$ Mpc/h, $s_{max}=70$ Mpc/h.
We limit the integral at both small and large scales.
At small scales the shape of $\xi_{\Delta s}(\mu)$ is seriously distorted by the finger of god (FoG) effect \citep{FoG}, 
and the distortion is more significant at lower redshift where structure undergoes more non-linear growth.
This introduces redshift evolution in $\xi_{\Delta s}(\mu)$ which is rather difficult to model.
As will be mentioned in \S 5, since FoG is particularly strong near the LOS direction,
in our analysis we also impose a cut $\mu>0.1$ to further suppress its contamination.
At large scales the measurement is dominated by noise due to poor statistics.
We find $s_{\rm min}=6-10$ Mpc/h and $s_{\rm max}=40-70$ Mpc/h are reasonable regions which can provide consistent results of 2pCF 
and tight, unbiased constraints on cosmological parameters.
In our analysis we choose $s_{\rm min}=6$ Mpc/h and $s_{\rm max}=50$ Mpc/h.

As an example Figure \ref{fig_Scatter} shows how the 2pCF is affected by AP and volume effects in the $\Omega_m=0.41$, $w=-1.3$ cosmology.
In choosing incorrect cosmological parameters, we expect the 2pCF to be influenced in three ways.
First, as a result of the AP effect, structures appear compressed in the radial direction.
This induces a nonuniform  variation in $\xi_{\Delta s}(\mu)$ as a function of angle.
Second, as a result of the volume effect, the sizes of structures shrink.
For example, a structure whose original size is $s_0=50$ Mpc/h will show up with a size $s_1<50$ Mpc/h.
As a result, the amplitude of $\xi_{\Delta s}(\mu)$ changes.
Finally, as another consequence of the volume effect, in the wrong cosmology structures on larger scales enter the statistics.
For example, halos within the blue solid box are not considered in the correct cosmology,
but they are taken into consideration in the wrong cosmology, as shown by the black dashed box.
This also results in a change in the amplitude of $\xi_{\Delta s}(\mu)$ since the binning in $s,\mu$-space will be 
inconsistent between different cosmological models. The combined effects of choosing an incorrect cosmology 
on shear and volume have been noted by \citet{topology}, who used the volume effects measured by the genus 
statistic to constrain the expansion history of the Universe.

% Note that the influence of volume effect can be more complicated with existence of AP or RSD effect.
% In that case, the volume effect may introduce extra anisotropy in the 2-point CF
% \footnote{
% At the radial and angular directions we have
% %\begin{equation}
% $Q_{\parallel} = \int_{s_{\rm min}}^{s_{\rm max}} ds\ \xi_{\parallel} (s),\ \ Q_{\perp} = \int_{s_{\rm min}}^{s_{\rm max}} ds\ \xi_{\perp} (s).$
% %\end{equation}
% Suppose the volume effect magnifies the length scale by a factor of $\tau$, i.e., 
% %\begin{equation}
% $s\rightarrow s^\prime= \tau s,\ $
% %\end{equation}
% This changes $\xi$ through a relation
% %\begin{equation}
% $\xi_\parallel(s) \rightarrow \xi^\prime_\parallel(s) = f_\parallel(\tau)\xi(s),
% \ \xi_\perp(s) \rightarrow \xi^\prime_\perp(s) = f_\perp(\tau)\xi(s).$
% %\end{equation}
% In the case that $f_\parallel(\tau) \neq f_\perp(\tau)$, 
% not only the values of $Q_{\parallel}$, $Q_{\perp}$ are changed,
% but also the relative ratio $Q_{\perp}/Q_{\parallel}$ is changed.
% This results in the extra anisotropy in the 2-point CF.
% In the case of no ARSD effect, we have $f_\parallel(\tau)=f_\perp(\tau) $ since $\xi$ is isotropic,
% but that is not guaranteed when there is RSD effect.
% }

%When measure $\int_{s_{\rm min}}^{s_{\rm max}} \xi (s,\mu_i)$ at different $\mu$ bins,
%not only the amplitude changes, but also the shape may change due to the anisotropic 

%In the upper panel of Figure \ref{}

%RSD will affect $\xi$, but the redshift dependence is small.
%{\it (put some formular)}
%We need to test that on simulations.

\section{Result}

%\subsection{Some Plottings, Discussions, Definition of $\chi^2$, ...}

%{\bf Figure: $\xi$ curves, in different cosmologies, without and with RSD}

\begin{figure*}
   \centering{
   \includegraphics[height=8cm]{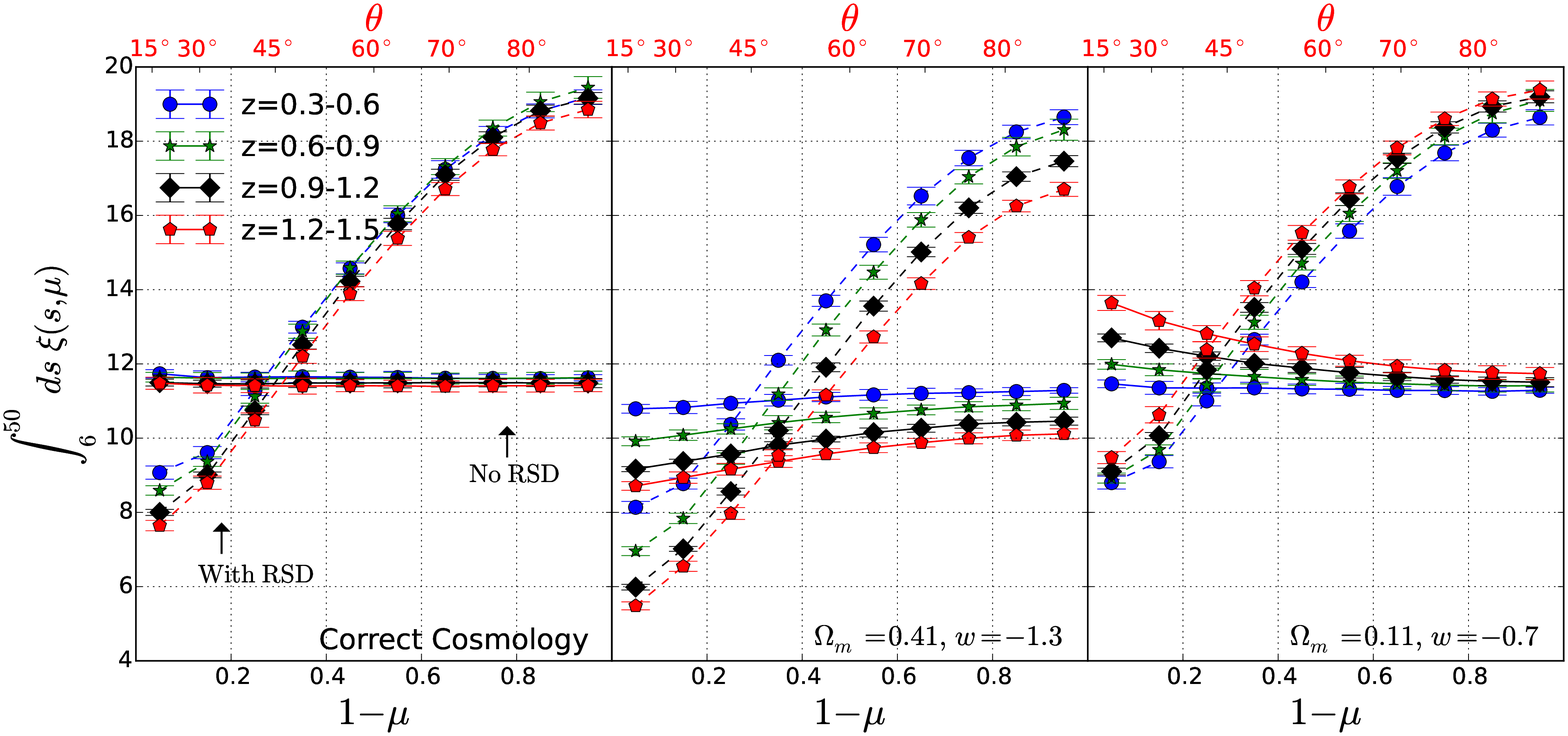}
   \includegraphics[height=8cm]{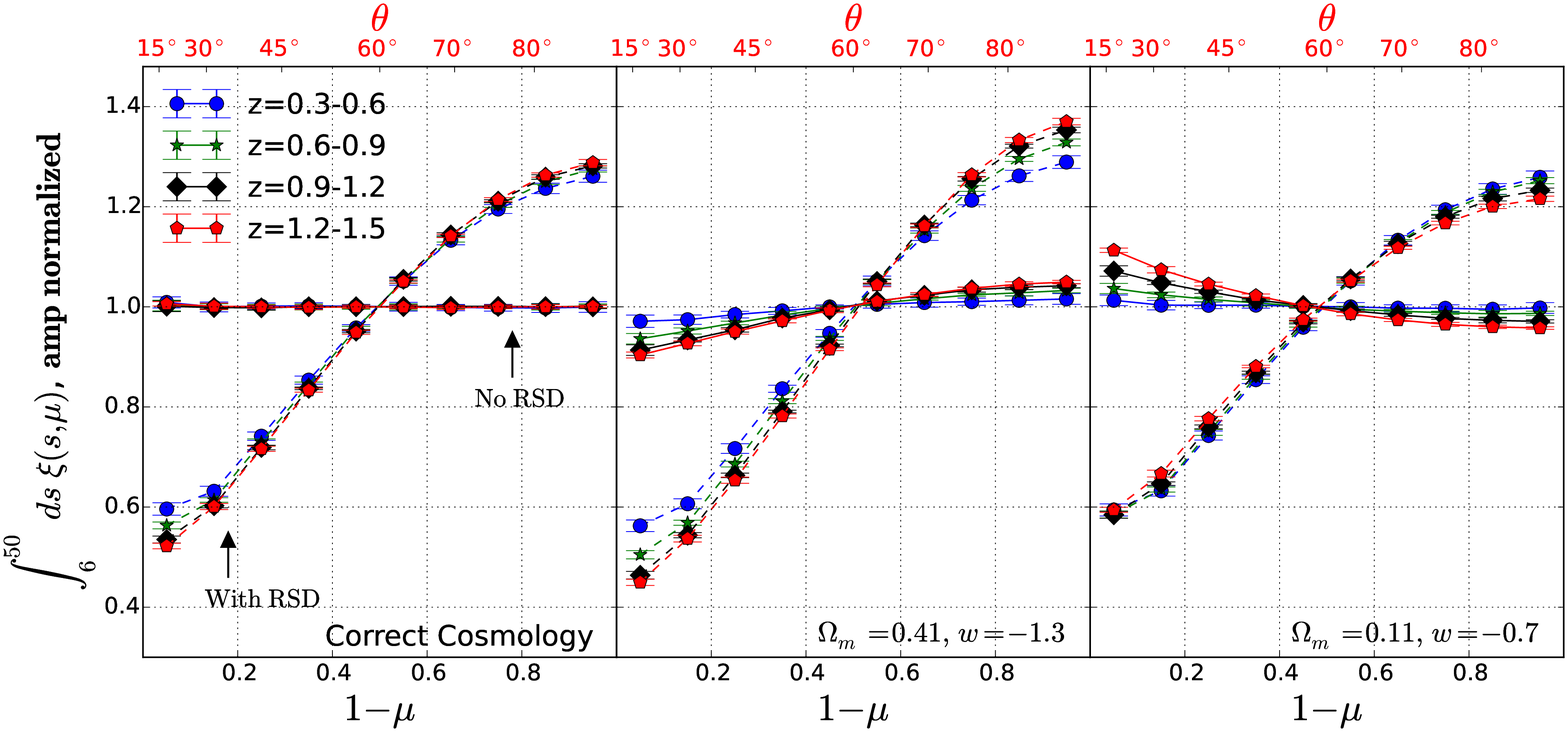}
   }
   \caption{\label{fig_TpCF}
   The 2pCF measured in four redshift bins, in the correct cosmology (left) and two wrongly assumed cosmologies (middle: $\Omega_m=0.11$, $w=-0.7$; 
   right: $\Omega_m=0.41$, $w=-1.3$). The clustering signal is measured as a function of $1-\mu$, where $\mu=\cos(\theta)$ and $\theta$ is the angle between the LOS and the vector joining the pair of galaxies. 
   Dashed and solid lines show the results with and without the RSD effect, respectively. 
   {\it Upper panel}: In the wrongly assumed cosmologies, we observe a clear change in the amplitudes and shapes of $\xi$ due to the volume and AP effect.
   Additionally, due to the redshift dependence of volume and AP effect,
   the amplitudes and shapes in the four redshift bins are different from each other. 
   {\it Lower panel}: The same as the upper panel, except that the amplitudes of curves are normalized to 1.
   }
\end{figure*}

In this section we present the measurements of 2pCF from our mock data.
Figure \ref{fig_TpCF} shows the $\xi_{\Delta s}$ measured from HR3 mock surveys,
adopting the correct cosmology (left), the $\Omega_m=0.11$, $w=-0.7$ cosmology (middle),
and the $\Omega_m=0.41$, $w=-1.3$ cosmology (right).
To study the redshift evolution, we divide the redshift range $z=0-1.5$ into five equal-width redshift bins
\footnote{We do not show the result of the first redshift bin, which is noisy due to poor statistics.}.
To measure anisotropy, we further divide the full angular range $\mu=0-1$ into 10 equal-width bins.
So we have %$\xi_{\Delta s}(z_i,\mu_j)$, where $i=1,2,...,5$, goes from 1 to 5, and $j$ goes from 1 to 10.
\begin{equation}
\xi_{\Delta s}(z_i,\mu_j)\ \equiv\ \xi_{\Delta s} \rm \ in\ the\ {\it i}-th\ redshift\ bin,\ {\it j}-th\ \mu\ bin
\end{equation}
where $i=1,2,...,5$, $j=1,2,...10$.
Measurements without and with considering RSD effect are plotted in solid and dashed lines, respectively.

The result for the correct cosmology is plotted in the upper left panel of Figure \ref{fig_TpCF}.
In the absence of RSD, we obtain flat curves of $\xi_{\Delta s}(\mu)$ in all redshift bins,
with the amplitude slightly different from one redshift bin to another.
This difference can arise from two sources; (a) The growth of clustering with the decreasing of redshift.
(b) The redshift evolution of the bias of halos having the same comoving density.
The result is significantly changed when we include the RSD effect.
Near the LOS direction ($\mu\rightarrow1$), structures are compressed due to the Kaiser effect,
so the value of $\xi_{\Delta s}(\mu)$ is smaller compared with measurements near the tangential 
direction\footnote{The FoG effect will enhance $\xi_{\Delta s}(\mu)$ in the LOS direction.
It does not significantly show up in our figures since we impose the cut $s_{\rm min}=6$ Mpc/h.}.
But it should be noted that the shape of $\xi_{\Delta s}(\mu)$ is nearly the same at all redshifts,
indicating the small redshift dependence of the RSD effect.
It is this observation that makes our method both feasible and statistically powerful. 
Even though the 2pCF becomes very anisotropic in redshift space, the anisotropy due to RSD does not change much as a function 
of redshift and its redshift-dependence is dominated by the geometric effects introduced by the adopted cosmology.

The results of the $\Omega_m=0.11$, $w=-0.7$ and $\Omega_m=0.41$, $w=-1.3$ cosmologies are plotted in the middle and right panels, respectively.
We can see that $\xi_{\Delta s}$ is significantly altered by the volume and AP effects.
In the $\Omega_m=0.41$, $w=-1.3$ cosmology,
the shrinkage of comoving volume suppresses the amplitude of $\xi_{\Delta s}(\mu)$,
and the LOS shape compression of structures results in a suppression of amplitude in the LOS direction compared with the tangential.
Both effects become increasingly more significant at higher redshift.
In the $\Omega_m=0.11$, $w=-0.7$ cosmology, the comoving volume shrinks at $z<0.6$ and enlarges at $z>0.6$.
Thus, compared to the result of correct cosmology (top left panel),
in this cosmology the 2pCF measured at $z=0.3-0.6$ has slightly suppressed amplitude,
while the other three curves measured at $z>0.6$ have enhanced amplitudes.
Also, due to the LOS shape stretch at $z>0.3$, in all curves the amplitude is relatively enhanced in the LOS direction.
All these effects are more significant at earlier times.

In real observational data the redshift evolution of the bias of observed galaxies is difficult to model.
Thus to mitigate this systematic uncertainty we wish to rely on the shape of $\xi_{\Delta s}(\mu)$, rather than its amplitude.
In the lower panel of Figure \ref{fig_TpCF}, we show the normalized $\xi_{\Delta s}(\mu)$ in each redshift bin, defined as
\begin{equation}\label{eq:norm}
 \hat\xi_{\Delta s}(\mu) \equiv \frac{\xi_{\Delta s}(\mu)}{\int_0^1\xi_{\Delta s}(\mu)\ d\mu}.
\end{equation}
It should be noted that the normalization in Equation (\ref{eq:norm}) is made by the correlation function
in each corresponding redshift bin, not by the global quantity. 
This procedure is critically important since it effectively suppresses the cosmic variance in the correlation function from one redshift bin to another.
When the RSD effect is considered,
in the correct cosmology $\hat {\xi}_{\Delta s}$ at different redshifts are almost identical to each other,
while in the wrong cosmologies we see clear redshift evolution.

Overall, the effect of RSD on the 2pCF is large but its redshift dependence is small.
The correct cosmology corresponds to the case with the lowest change of $\xi_{\Delta s}$ with $z$.
Even with RSD, we can still correctly determine the true cosmology by using the relative change of $\xi_{\Delta s}$ with redshift.
Based on this fact, we define our $\chi^2$ as follows
\begin{equation}\label{eq:chisq1}
\chi^2\equiv \sum_{i=1}^{4} \sum_{j=1}^{10} \frac{[\hat\xi_{\Delta s}(z_i,\mu_j)-\hat\xi_{\Delta s}(z_5,\mu_j)]^2}
{\sigma_{\hat\xi_{\Delta s}(z_i,\mu_j)}^2+\sigma^2_{\hat\xi_{\Delta s}(z_5,\mu_j)}}.
\end{equation}
The 2pCF measured in the 1-4 redshift bins are compared (or normalized) to the measurement in the last redshift bin.
This $\chi^2$ will prefer minimal shape change over the redshift range, with little of no weight given to the amplitude of the clustering statistic. 

Figure \ref{fig_TpCF} shows that the redshift evolution of the RSD effect, although small,
still results in visible redshift evolution in the amplitudes and shapes of the curves.
Also, the amplitude of the 2pCF may evolve due to the growth of clustering and the redshift evolution of selection effect.
So, similar to \cite{Li2014}, we further correct the residual RSD effect, i.e.,
the following quantity is computed in the correct cosmology and subtracted from our results,
\begin{equation}\label{eq:rsdcor}
 \Delta \hat\xi_{\Delta s} \equiv \hat\xi_{\Delta s, {\rm With\ RSD}} - \hat\xi_{\Delta s, \rm No\ RSD}.
\end{equation}
Although this correction factor may have cosmological dependence, we assume that this will not introduce significant systematic to our methodology.
To avoid the contamination from the FoG effect we further impose a cut of $\mu>0.1$ (i.e. $j>1$).
We also take into consideration the correlation between $\xi_{\Delta}(\mu)$ measured in different directions.
So the $\chi^2$ function becomes
\begin{equation}\label{eq:chisq1}
\chi^2\equiv \sum_{i=1}^{4} \sum_{j_1=2}^{10} \sum_{j_2=2}^{10} {\bf p}(z_i,\mu_{j_1}) {\bf Cov}_{i,j_1,j_2}  {\bf p}(z_i,\mu_{j_2}),
\end{equation}
where ${\bf p}(z_i,\mu_{j})$ takes the form
\begin{eqnarray}\label{eq:bfp}
 {\bf p}(z_i,\mu_{j}) \equiv&\ \hat\xi_{\Delta s}(z_i,\mu_j)-\hat\xi_{\Delta s}(z_5,\mu_j) - \nonumber\\
   &\ \ \ \ \ [\Delta\hat\xi_{\Delta s}(z_i,\mu_j)-\Delta\hat\xi_{\Delta s}(z_5,\mu_j)],
\end{eqnarray}
and ${\rm Cov}_i$ is the covariance matrix of ${\bf p}(z_i,\mu_{j})$. 
In testing our methodology, we divide the full sky sample into 48 equal sky coverage subsamples
and construct a measurement covariance matrix. 
%from the 48 equal sky-area mock surveys described in \S\ref{sec:mocks}

\begin{figure*}
   \centering{
   \includegraphics[height=7cm]{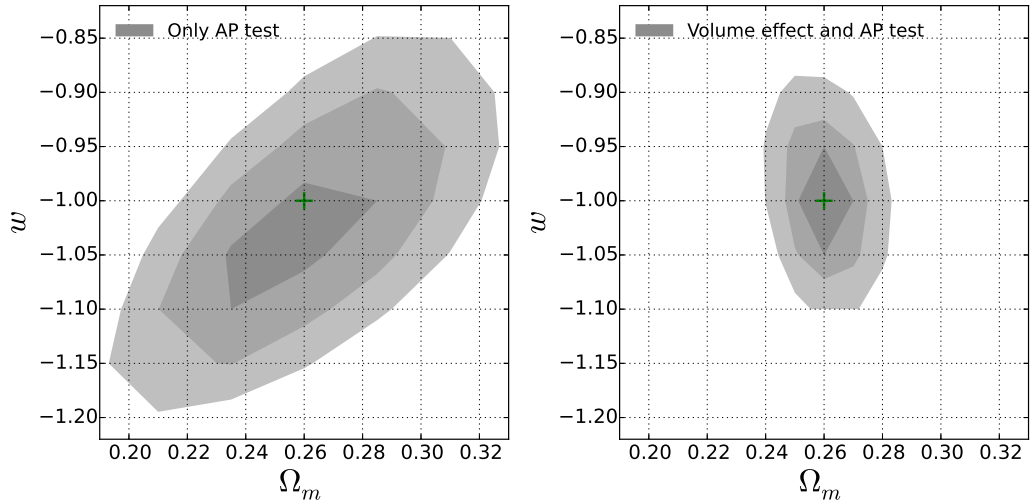}
   }
   \caption{\label{fig_contours}
   {\em Left:} Expected likelihood contours (68.3\%, 95.4\%, and 99.7\%) in the $\Omega_m-w$ plane,
   obtained from a $1/8$-sky, $z<1.5$ mock survey with a constant galaxy number density of $\bar n=3\times10^{-4}(h^{-1}{\rm Mpc})^{-3}$.
   We achieve unbiased constraints with $\delta w\sim0.1$ and $\delta \Omega_m\sim0.03$ 
   by comparing the shapes of $\xi_{\Delta s}(\mu)$ measured in different redshift bins.
   {\em Right:} Here we use the unnormalized $\xi_{\Delta s}(\mu)$, 
   which is sensitive to the volume change and thus provides much tighter constraints. 
   Although to use this in practice would mean overcoming some observational systematic uncertainties like galaxy evolution and selection bias.
   }
\end{figure*}

Figure \ref{fig_contours} shows the likelihood contours obtained from our mock survey, 
which is a $1/8$-sky survey with constant number density $\bar n=3\times10^{-4}(h^{-1}{\rm Mpc})^{-3}$ in the redshift range 0-1.5.
The gray contours on the left show the result when we normalize the amplitude of $\xi_{\Delta s}(\mu)$ 
and just utilize the information encoded in the shape
(i.e., we focus only on the AP information and ignore the volume effect).
We obtain tight and unbiased constraint on $\Omega_m$ and $w$, with 68.3\% CL intervals $\delta \Omega_m\sim0.03$ and $\delta w\sim0.1$.
We find that $\Omega_m$ and $w$ are positively degenerated with each other, which is expectable.
For instance, reducing $\Omega_m$ and having a more phantom-like dark energy produce similar influences on the expansion history of the Universe. 
Compared with the result of our gradient field method \cite{Li2014},
the shapes of contours are very similar, and constraint is a little tighter.

The right hand plot of Figure \ref{fig_contours} shows the result when we utilize the volume effect.
To do that, in Equation (\ref{eq:bfp}) we adopt the non-normalized $\xi_{\Delta s}$ instead of the normalized one $\hat\xi_{\Delta s}$
\footnote{Different from the normalized case, in the non-normalized case not only do we correct the RSD effect, 
but we also correct the redshift evolution of the amplitude of 2pCF.
As shown by the top left pannel of Figure \ref{fig_TpCF}, the amplitude of 2pCF can evolve with redshift even in the no RSD case,
mainly due to the growth of structure and the selection bias.
So in the non-normalized case Equation (\ref{eq:rsdcor}) shall be modified as  
$\Delta \xi_{\Delta s}(z_i) \equiv \left[\xi_{\Delta s, {\rm With\ RSD}}(z_i) - \xi_{\Delta s, \rm No\ RSD}(z_i)\right]
+ \left[\xi_{\Delta s, \rm No\ RSD}(z_i) - \xi_{\Delta s, \rm No\ RSD}(z_1)\right]$.}. 
In that case, constraints become much tighter, with $\delta \Omega_m\sim0.007$ and $\delta w\sim0.035$.
Also, the direction of degeneracy changes and is very different from mainstream techniques of CMB, SNIa and BAO,
meaning that combining our method with these techniques can significantly improve the constraint.
To implement it in real observational cases, it is necessary to model the evolution of the clustering amplitude for the observed galaxies.

\section{Conclusion}

\begin{figure*}
   \centering{
   \includegraphics[height=7cm]{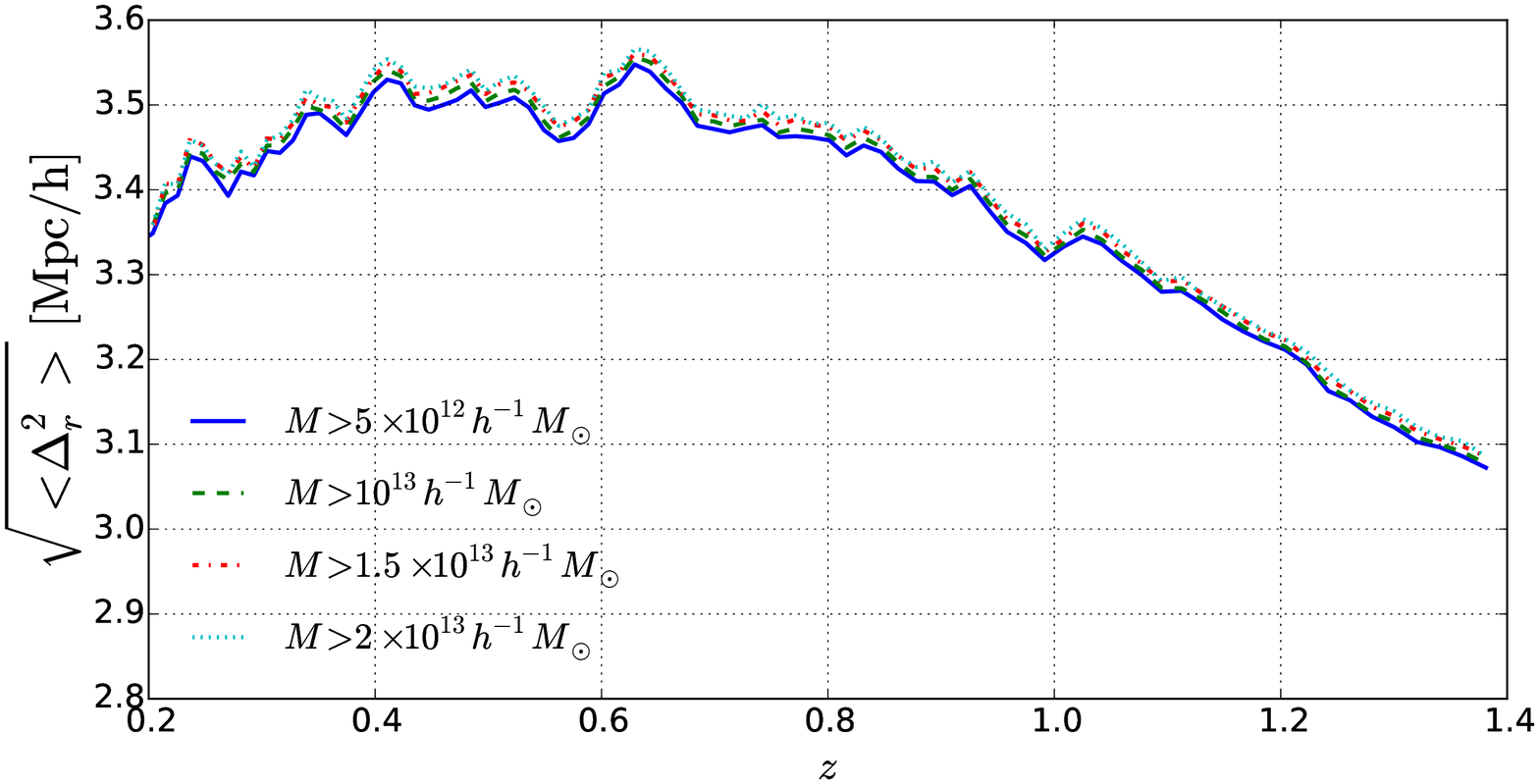}
   }
   \caption{\label{fig_deltar}
   The RMS displacement of halo comoving distance as a function of redshift, for samples with different cuts of halo mass.
   %The RMS peculiar velocity divided by $H(z)$ as a function of radial diastase from the observer (equivalently to redshift) for various mass cuts. 
   Over the redshift range of interest for our study, the peculiar velocity is varying smoothly with a maximal deviation of 13\%.
   Within the wide mass cut range from $5\times 10^{12}$\hMsun to $2\times 10^{13}$\hMsun, 
   the difference in the RMS displacement is $\lesssim1\%$,
   suggesting that the redshift evolution of RSD effect shall be fairly insensitive to the halo mass cut and number density of the sample.
   }
\end{figure*}

We have presented a new anisotropic clustering statistic that can probe the cosmic expansion history.
We measure the integrated 2pCF, $\xi_{\Delta s}(\mu)\equiv\int_{s_{\rm min}}^{s_{\rm max}}\xi(s,\mu)ds$, as a function of direction $\mu$.
The amplitude of $\xi_{\Delta s}(\mu)$ is affected by the volume effect,
and the shape is affected by the AP effect.
Due to the redshift dependence of the volume and AP effects, 
in wrongly adopted cosmologies there are redshift evolutions of the amplitude and shape.
The RSD effect due to galaxy peculiar velocities, although having a strong effect on $\xi_{\Delta s}(\mu)$,
does not exhibit significant redshift evolution.
Thus by focusing on the redshift dependence of $\xi_{\Delta s}(\mu)$,
we are able to derive accurate and unbiased estimates of cosmological parameters in spite of contamination induced by RSD.
We are only assuming that the RSD modelling is accurately captured by the simulations.

The concept of this paper is similar to \cite{Li2014}, 
where the redshift dependence of the AP effect is measured from the anisotropy in the galaxy density gradient field.
However, in this paper we choose a different statistical method, i.e. the 2pCF.
%Both the methods of density gradient field and 2pCF can characterize the anisotropy of the galaxy distribution,
They differ from each other in several aspects.
1) Using the 2pCF method it is more convenient to choose the scales we investigate.
2) The advantage of the density gradient field method is that, it allows us to utilize the information on small scales of $\sim$10 Mpc/h 
(depending on the scale of smoothing).
3) In the 2pCF method we are able to probe the volume effect, which is not possible for the galaxy density field method.
4) The 2pCF is a mature statistic in cosmology and its optimal estimation and statistical properties are well understood.

The volume effect, which causes redshift evolution in the amplitude of 2pCF,
leads to very tight constraint on cosmological parameters.
But it suffers from systematic effects of growth of clustering and the variation of galaxy sample with redshift.
It would be great if one can reliably model these two effects and utilize the volume effect.
In case that the systematic effect can not be correctly modelled,
one can focus on the AP effect by normalizing the amplitude of $\xi_{\Delta s}(\mu)$ and just investigating the redshift evolution of the shape.

In our analysis we consider the case of a flat universe with constant dark energy EoS $w$. 
However, in general one can choose an arbitrary cosmological model and repeat our analysis to constrain the model parameters. 
Furthermore, since our method focuses on the redshift evolution of the 2pCF, 
it may have advantages in constraining models in which the cosmic expansion history displays dynamical features. 
For the popular Chevallier-Polarski-Linder (CPL) parametrization $w=w_0+w_a \frac{z}{1+z}$ \citep{CP2001,Linder2003}, 
the dark energy EoS evolves monotonically from $w_0$ at $z=0$ to $w_0+w_a$ at high redshift. 
Using a LSS sample covering a wide redshift range should be sufficient to constrain the model parameters through the effect on $D_A$ and $H$. 
For dark energy models with a rapidly varying EoS \citep{Bassett2002,Corasaniti2003}, 
a transition of $w$ could happen in a very short period, which is difficult to be detected by common techniques. 
For our method, we can reduce the redshift bin size down the a size similar to the typical width of a transition epoch 
to probe the detailed feature of cosmic expansion history and effectively constrain the model parameters.

The feasibility of our method crucially depends on the fact that the redshift dependence of the RSD effect is small. 
One may suspect that this is a special consequence of the constant number density sample. 
While an exhaustive investigation of this issue is too complicate for this short paper, 
we perform a simple check to show that the weak redshift dependence of the RSD effects is fairly universal. 
From the HR3 lightcone data, we generated a series of samples with a constant mass cut, 
and measured the root mean square (RMS) displacement of halo comoving distance as a function of redshift. 
The degree of the RSD effects shall be statistically directly proportional to the RMS displacement. 
The result is plotted in Figure \ref{fig_deltar}.
We find that the total variation of the RMS displacement between $z=0$ and 1.5 is only about 13\%, which can explain why the redshift dependence of the RSD effects is weak.
We also find that, within the wide mass cut range from $5\times 10^{12}$\hMsun to $2\times 10^{13}$\hMsun, the difference in the RMS displacement is $\lesssim1\%$. 
This suggests that the redshift evolution of RSD effect shall be fairly insensitive to the halo mass cut and number density of the sample.

When dealing with real observational data, 
it will be important to accurately model the galaxy clustering to remove the residual RSD effects on the 2pCF.
It will also require the handling of various observation effects such as survey geometry, fiber collisions, etc. 
We will report the results of such investigations in forthcoming studies.

\section*{Acknowledgments}

We thank the Korea Institute for Advanced Study for providing computing resources (KIAS Center for Advanced Computation Linux Cluster System).
We thank Seokcheon Lee and Graziano Rossi for many helpful discussions.

\bsp

\label{lastpage}

\end{document}